\documentclass[twocolumn,showpacs,preprintnumbers,amsmath,amssymb]{revtex4}
\usepackage[utf8]{inputenc}
\usepackage[english]{babel}
\usepackage{graphicx}
\usepackage{dcolumn}
\usepackage{bm}

\begin{document}

\title{Change in the Orbital Period of a Binary System Due to Dynamical Tides
for Main-Sequence Stars}
\author{S.V. Chernov}
\affiliation{\it Astrospace Center, Lebedev Physical Institute, Russian Academy of Sciences, Profsoyuznaya ul. 84/32,
Moscow, 117997 Russia}
\email{chernov@lpi.ru}

\begin{abstract}
We investigate the change in the orbital period of a binary system due to dynamical tides by
taking into account the evolution of a main-sequence star. Three stars with masses of one, one and a half, and two solar masses are considered. A star of one solar mass at lifetimes $t=4.57\times10^9$ yr closely corresponds to our Sun. We show that a planet of one Jupiter mass revolving around a star of one solar mass will fall onto the star in the main-sequence lifetime of the star due to dynamical tides if the initial orbital period of the planet is less than $P_{\rm orb}\approx2.8$ days. Planets of one Jupiter mass with an orbital period$P_{\rm orb}\approx2$ days or shorter will fall onto a star of one and a half and two solar masses in the mainsequence
lifetime of the star.
\end{abstract}
\maketitle
\section{Introduction}

Tidal interactions play an important role in dynamical
processes in the two-body problem in close
binary systems: star–star (binary star) or star–
planet. They can lead to such phenomena as synchronization
and orbital circularization (Hut 1981;
Zahn 1977) as well as to the tidal capture (Press
and Teukolsky 1977) or disruption of an object (star)
(Ivanov and Novikov 2001) and the fall of the object
onto the star (Rasio et al. 1996; Penev et al. 2012;
Bolmont and Mathis 2016). In this paper we consider
the tidal interaction of two bodies: a star and a point
source. The point source can be both a star (a neutron
star, a white dwarf, etc.) and a planet. Below we
will call the point source a planet by implying that
this can also be a star. Since the evolution time
scales of the eccentricity or semimajor axis strongly
depend on the orbital period of the binary system
and for some stellar models can take values up to
108 yr or more for periods of about 5 days (Ivanov
et al. 2013; Chernov et al. 2013), the evolution of the
star itself should be taken into account on such time
scales. As the star evolves, the orbital parameters
change due to tidal interactions. In this paper we
investigate the dynamical tides by taking into account
the stellar evolution. For our study we chose three
types of stars with masses of one, one and a half,
and two solar masses. We consider all stars without
allowance for their rotation and magnetic field and
touch on the stellar physics itself superficially, as far
as this problem requires. The star of one solar mass
at lifetimes $t = 4.57\times10^9$ yr closely corresponds to
our Sun and has a radiative core and a convective
envelope on the main sequence. The other two stars
of one and a half and two solar masses are more
massive and have a more complex structure. These
stars have a convective core and a radiative envelope
on the main sequence (a more precise structure is
presented below).

The problem of determining the tidal evolution is
reduced to the problem of determining the normal
modes of stellar perturbations and to calculating
the energy and angular momentum exchange in the
star.planet system (Ivanov and Papaloizou 2004,
2010; Papaloizou and Ivanov 2010; Lanza and
Mathis 2016). The low-frequency g-modes of the
stellar oscillations play an important role in the
theory of dynamical tides. The tidal interactions
are fairly intense at small periastron distances of the
planet. For a periastron distance $a\approx 0.01$ AU, the
dimensionless excitation frequency is $\tilde{\omega}\sim0.3$, which
corresponds to g-modes.

A large number of exoplanets in stellar systems
have been discovered in the last few years owing
to the Kepler, SuperWasp, and other observational
programs. In particular, short-period massive planets
with an orbital period of a few days, the so-called hot
Jupiters, have been detected (Winn 2015). As a rule,
the hot Jupiters have low eccentricities, which points
to the importance of tidal interactions (Ogilvie 2014).
The results of this paper can be directly applied to
some of such systems. For example, the system
YBP1194 is a solar twin (Brucalassi et al. 2014).
A planet with a mass of 0.34$M_J$ and a period of
only 6.9 days revolves around this star. For such
a short-period planet the dynamical tides must be
fairly intense and must affect the orbital evolution.
Predictions about the subsequent evolution of this
planet can be made by analyzing this system.

One of the results of this evolution is the fall of
the planet onto the star. The possibility of such a fall
has been considered in many papers (see, e.g., Rasio
et al. 1996; Penev et al. 2012; Weinberg et al. 2012;
Essick and Weinberg 2016). Rasio et al. (1996)
considered the possibility of the fall of the planet
onto the star due to quasi-static tides and provided
a plot for solar-like stars that shows the threshold,
as a function of planetary mass and orbital period,
below which the planet falls onto the star. Penev
et al. (2012) considered tides with a constant tidal
quality factor $Q^{'}$ specified phenomenologically. In
reality, this factor will depend on the planet’s orbital
period (Ivanov et al. 2013) and stellar age. Essick
and Weinberg (2016) took into account the energy
dissipation due to nonlinear interaction of modes with
one another. In contrast to our approach (Ivanov
et al. 2013; Chernov et al. (2013), the simultaneous
solution of a large number of ordinary differential
equations for each stellar model is suggested, with
only solar-type stars having been considered.

In this paper we consider the evolution of stars
with masses of one, one and a half, and two solar
masses. Data on the stars are presented in Tables 1–
3. A novelty of this study is a consistent allowance for
the stellar evolution. For each moment of the star’s
lifetime we calculated the spectra of normal modes,
the overlap integrals (Press and Teukolsky 1977),
which are a measure of the intensity of tidal interactions
(for a generalization to the case of a rotating
star, see Papaloizou and Ivanov 2005; Ivanov and
Papaloizou 2007), and the time scales of the orbital
parameters (the tidal quality factor $Q^{'}$). The overlap
integrals are directly related to the tidal resonance
coefficients that were introduced by Cowling (1941)
(see also Rocca 1987; Zahn 1970).

All of the quantities marked by a tilde are dimensionless;
the normalization is presented in Appendix
A.

\section{FORMULATION OF THE PROBLEM}

Such a quantity as the overlap integral Q is of
great importance in the theory of dynamical tides.
It is specified by the expression (Press and Teukolsky
1977; Zahn 1970)
\begin{eqnarray}
 \tilde{Q} &=& \sqrt{\frac{3}{4\pi}}
 \frac{\int\limits_0^1 x^{l+1}\tilde{\rho}l(\xi_r+(l+1)\xi_s)dx}
 {\sqrt{\int\limits_0^1 x^2\tilde{\rho}(\xi_r^2+l(l+1)\xi_s^2)dx}},\nonumber\\
 Q &=& \tilde{Q}\sqrt{M}R,
 \label{OverlapsIntegral}
\end{eqnarray}
where M is the stellar mass, R is the stellar radius,
and $\tilde{Q}$ is the dimensionless overlap integral; the remaining
quantities and their dimensionless forms are
defined in Appendix A. The overlap integral serves as
a measure of the intensity of the tidal forces and plays
a crucial role in the physics of dynamical tides (Press
and Teukolsky 1977; Zahn 1970, 1975, 1977). Here,
we will consider only the quadrupole part of the tidal
forces $l = 2$, because it is this part that makes the
greatest contribution to the overlap integral (Press
and Teukolsky 1977). The change in orbital semimajor
axis and eccentricity with time is determined
from the energy and angularmomentum conservation
laws and is specified by the following formulas (Ivanov
et al. 2013):
\begin{eqnarray}
 \frac{\dot{a}}{a}=-\frac{2}{T_a},\qquad\frac{\dot{e}}{e}=-\frac{1}{T_e},
 \label{TaTe}
\end{eqnarray}
Here, $e$ is the orbital eccentricity, $a$ is the semimajor
axis, and the time scales of the change in semimajor
axis $T_a$ and eccentricity $T_e$ in the case of low eccentricities
($e\rightarrow0$) are specified by the relations (Ivanov
et al. 2013)
\begin{eqnarray}
 T_a=-\frac{GM M_2}{a\dot{E}_I},\quad
 T_e=-\frac{GM M_2e^2}{a\dot{E}_e},
 \label{TaTe1}
\end{eqnarray}
where $M_2$ is the planetary mass. The rate of change
of energy $\dot{E}_I$, $\dot{E}_e$ in the case of a dense spectrum and
moderately large viscosities is specified by Eqs. (51)
from Ivanov et al. (2013). The spectrum is deemed to
be dense enough if $\frac{d\tilde{\omega}}{dj}<<\tilde{\omega}$. For simplicity, we will
consider here only the time scales of the changes in
semimajor axis. Using formulas from Subsection 7.1
of the paper by Ivanov et al. (2013), we obtain
\begin{eqnarray}
 T_a=\frac{10}{3\pi^2}\bigg|\frac{d\tilde{\omega}}{dj}\bigg|\frac{(1+\mu)^{5/3}}{\mu \tilde{Q}^2\Omega_\ast}
 \left(\frac{P_{\rm orb}\Omega_\ast}{2\pi}\right)^{10/3},
 \label{Ta}
\end{eqnarray}
where $\mu=\frac{M_2}{M}$ is the ratio of the planetary mass
to the stellar mass, $\frac{d\tilde{\omega}}{dj}$ is the difference between
adjacent frequencies, $j$ is the frequency number, and
$\Omega_\ast=\sqrt{\frac{GM}{R^3}}$ (Ivanov et al. 2013). The orbital
period of the planet around the star is specified by the
relation
\begin{eqnarray}
 P_{\rm orb}=2\pi\sqrt{\frac{a^3}{G(M+M_2)}}.
 \label{Porb}
\end{eqnarray}
Below, using Eqs. (\ref{Ta}) and (\ref{Porb}) and solving Eq. (\ref{TaTe}), we
will present the change in the planet’s orbital period
as a function of time by taking into account the stellar
evolution.

As is clear from Eq. (\ref{OverlapsIntegral}), the eigenfunctions and,
accordingly, eigenfrequencies of the stellar oscillations
should be known to calculate the overlap integral.
There exist three types of eigenfrequencies for a
nonrotating star: the so-called p-, f-, and g-modes.
We will be concerned only with the low-frequency g- modes.
The properties of the g-modes are determined
by the Brunt - V$\ddot{\rm a}$is$\ddot {\rm a}$l$\ddot {\rm a}$ frequency. It is specified as
follows (Christensen-Dalsgaard 1998):
\begin{eqnarray}
 N^2=g\left(\frac{1}{\Gamma p}\frac{\partial p}{\partial r}-
 \frac{1}{\rho}\frac{\partial\rho}{\partial r}\right),
\end{eqnarray}
where $p$ is the pressure, $\rho$ is the density, $g$ is the
gravitational acceleration, $\Gamma=(\partial\ln p/\partial\ln\rho)_{\rm ad}$ is
the adiabatic index. However, this definition of the
Brunt - V$\ddot{\rm a}$is$\ddot {\rm a}$l$\ddot {\rm a}$ frequency is difficult to apply to many
stars. This is related to the numerical errors, to the
calculation of the derivative of the density, and to
the fact that there exist regions in stars where both
terms in parentheses can be of the same order of
magnitude. Therefore, a different definition is used
(Brassard et al. 1991):
\begin{eqnarray}
 N^2=\frac{g^2\rho}{P}\frac{\chi_{T}}{\chi_\rho}\bigg[\nabla_{\rm ad}-\nabla-
 \frac{1}{\chi_T}\sum_{i=1}^{N-1}\chi_{X_i}\frac{d \ln X_i}{d\ln p}\bigg],
\end{eqnarray}
where
\begin{eqnarray}
 \nabla_{\rm ad}=\left(\frac{\partial\ln T}{\partial\ln P}\right)_{{\rm ad},X_i},\quad
 \chi_{T}=\left(\frac{\partial\ln p}{\partial\ln T}\right)_{\rho,X_i},\nonumber\\
 \chi_\rho=\left(\frac{\partial\ln p}{\partial\ln\rho}\right)_{T,X_i},\quad
 \chi_{X_i}=\left(\frac{\partial\ln p}{\partial\ln X_i}\right)_{\rho,T,X_{i\ne j}},
\end{eqnarray}
the temperature gradient
\begin{eqnarray}
  \nabla=\frac{\partial\ln T}{\partial\ln P},
\end{eqnarray}
$X_i$ is the mass fraction of atoms of type i, and
\begin{eqnarray}
 \sum_{i=1}^{N-1}X_i+X_N=1.
\end{eqnarray}
It is convenient to divide the Brunt - V$\ddot{\rm a}$is$\ddot {\rm a}$l$\ddot {\rm a}$ frequency
into two parts: the structural part (structure terms)
Brunt - V$\ddot{\rm a}$is$\ddot {\rm a}$l$\ddot {\rm a}$
\begin{eqnarray}
 N^2_{\rm st}=\frac{g^2\rho}{P}\frac{\chi_{T}}{\chi_\rho}\bigg[\nabla_{\rm ad}-\nabla\bigg]
\end{eqnarray}
and the compositional part
\begin{eqnarray}
 N^2_{\rm com}=-\frac{g^2\rho}{P}\frac{1}{\chi_\rho}
 \sum_{i=1}^{N-1}\chi_{X_i}\frac{d \ln X_i}{d\ln p}.
\end{eqnarray}
The compositional part is related to the change in
stellar chemical composition due to nuclear reactions,
and it is of great importance at the boundary of the
convective and radiative regions. Another important
stellar characteristic is the acoustic frequency defined
by the formula
\begin{eqnarray}
 S^2_l=\frac{l(l+1)c^2}{r^2}.
\end{eqnarray}

The low-frequency g-modes are excited at $\omega^2<N^2$.
The high-frequency p-modes are excited at $\omega^2>S^2_l$.
We will not consider the p-modes, because they play
a secondary role in the theory of dynamical tides.

Thus, the problem is reduced to finding the eigenfunctions
and eigenfrequencies of the stellar oscillations,
calculating the overlap integrals, and calculating
the characteristic orbital parameters as a
function of time. The equations that describe the
eigenfrequencies and eigenfunctions of a star in the
adiabatic and Cowling approximations are presented
in Appendix A. The stars were modeled with the
MESA software package (Paxton et al. 2011, 2013,
2015); the data obtained were then interpolated by the
Steffen (1990) method to two million points. Using
these data, we solved the eigenfrequency and eigenfunction
problem and calculated the overlap integrals.
The fourth-order Runge–Kutta method was used to
solve the differential equations. The data for the stars
and the overlap integrals were compared with those
from Ivanov et al. (2013) and Chernov et al. (2013).

\section{RESULTS}

In this section we present the results of our calculations
of the change in the planet’s orbital period
for three types of stars with masses of one, one and a
half, and two solar masses. The course of stellar evolution
for these models is shown on the Hertzsprung–
Russell diagram depicted in Fig. 1. Data for each star
are presented in the tables in Appendix B. The main
phase of stellar evolution is the main-sequence phase.
It is at this phase that any star spends most of its
lifetime. This phase is indicated in Fig. 1 by the nearly
horizontal straight line.

\begin{figure}
\includegraphics[width=0.46\textwidth]{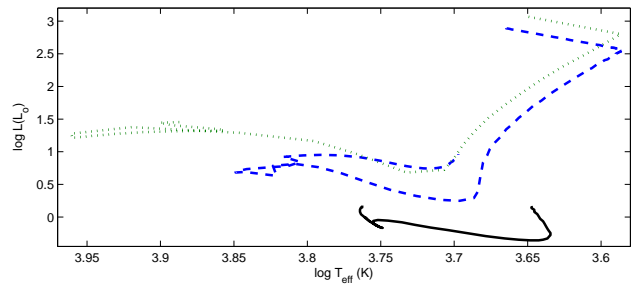}
\caption{Hertzsprung–Russell diagram for three stars: the solid, dashed, and dotted lines are for the Sun, the star
of one and a half solar masses, and the star of two solar masses, respectively.}
\label{HR}
\end{figure}

\subsection{The Sun}

In this section we present the results of our calculations
of the overlap integrals for various lifetimes
of the star of one solar mass and the change in the
planet’s orbital period with time. The star was modeled
from $t=1.46\times 10^6$ to $t=8.61\times 10^9$ yr. At times
$t=4.57 \times 10^9$ yr this star closely corresponds to our
Sun, and below we will call this star the Sun. At
the initial time the Sun was completely convective,
hydrogen burning has not yet begun, and its metallicity
is z = 0.02. Figure 2 shows a dimensionless
dependence of the solar density on radius at the initial
and subsequent times.

\begin{figure}
\includegraphics[width=0.45\textwidth]{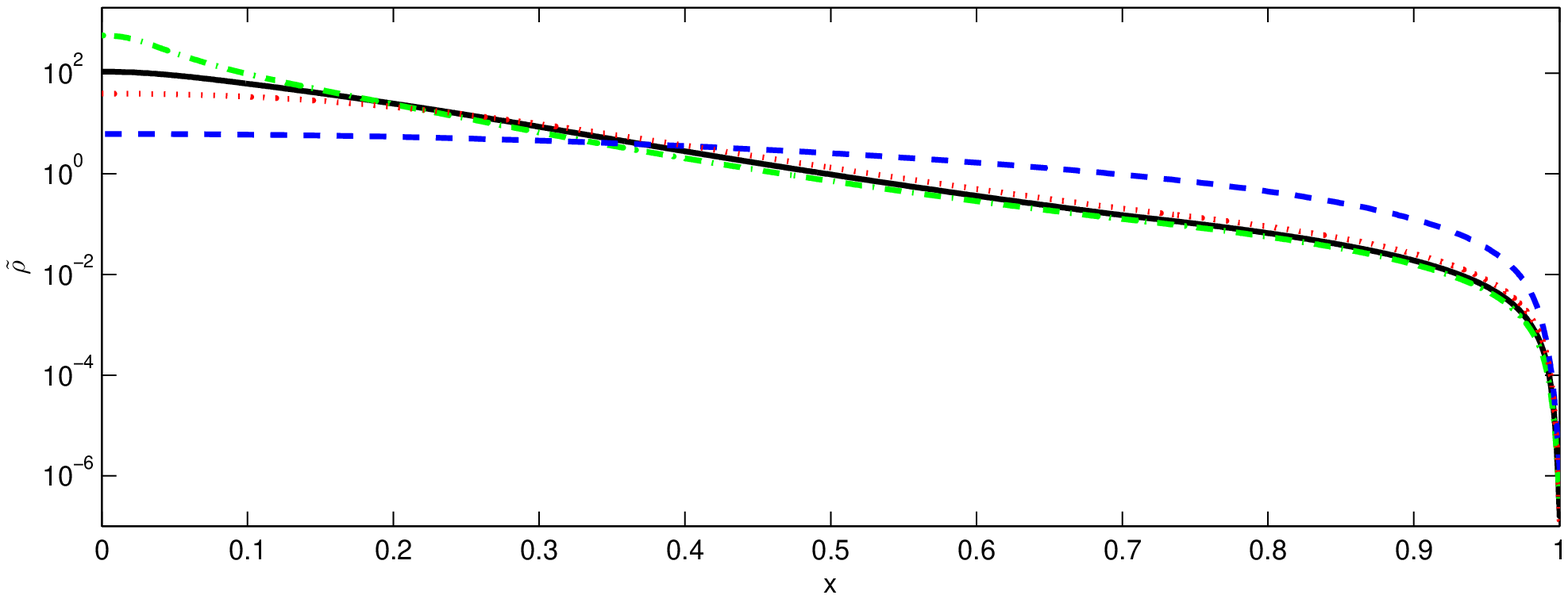}
\includegraphics[width=0.45\textwidth]{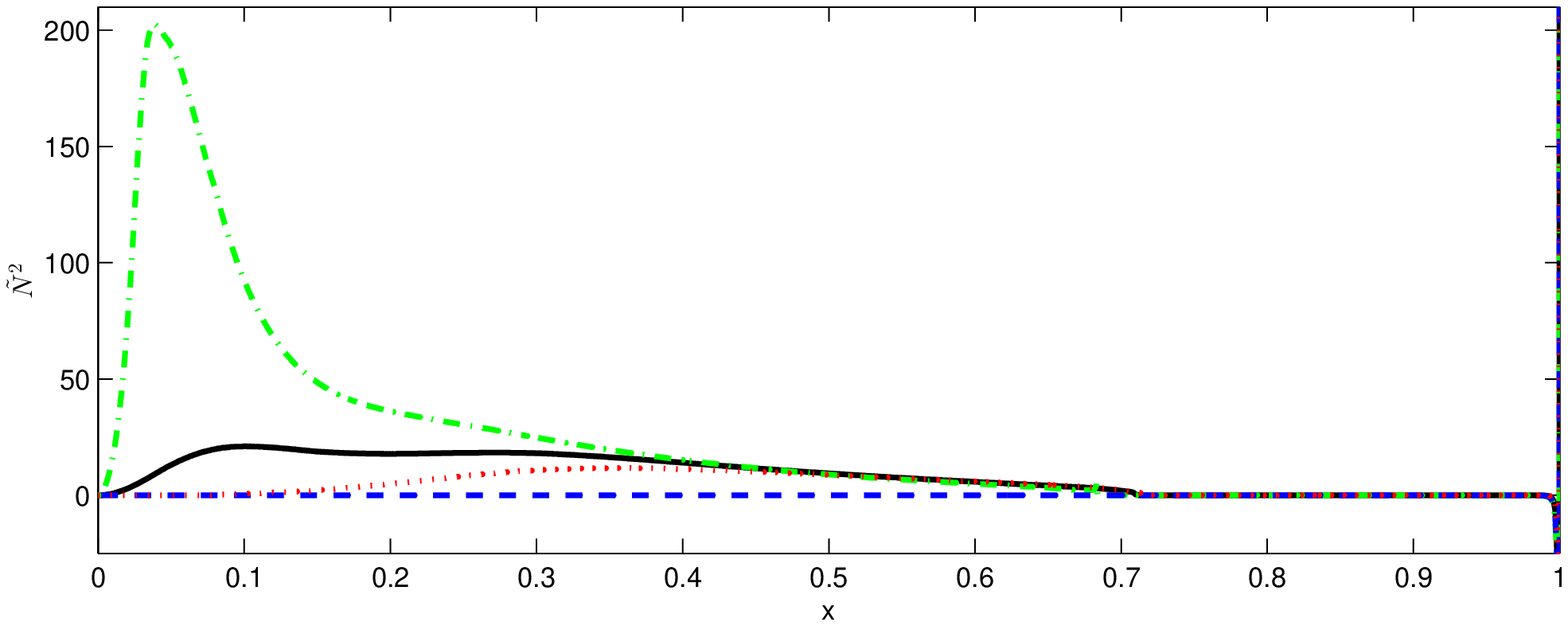}
\caption{Density (a) and Brunt - V$\ddot{\rm a}$is$\ddot {\rm a}$l$\ddot {\rm a}$ frequency (b) for the Sun versus radius for various times: the dashed,
dotted, solid, and dash–dotted lines are for the ages $t=1.46\times 10^6$, $t=1.02\times 10^8$, $t=4.57\times 10^9$, and $t=8.61\times 10^9$ yr, respectively.}
\label{fig1}
\end{figure}

\begin{figure}
\includegraphics[width=0.45\textwidth]{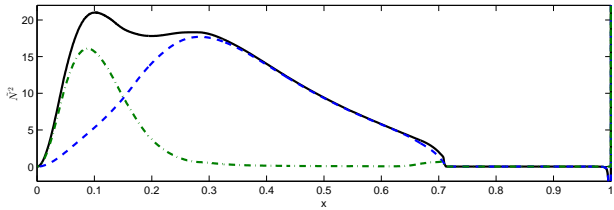}
\caption{Brunt - V$\ddot{\rm a}$is$\ddot {\rm a}$l$\ddot {\rm a}$ frequency versus radius for the solar age $t=4.57\times 10^9$: the solid, dashed, and
dash-dotted lines indicate the total Brunt - V$\ddot{\rm a}$is$\ddot {\rm a}$l$\ddot {\rm a}$ frequency $N^2$, the structural part $N^2_{\rm st}$, and the compositional part $N^2_{\rm com}$,
respectively.}
\label{fig2}
\end{figure}

The course of stellar evolution and the excitation
of eigenmodes can be described as follows. At times
$t\approx1.46\times 10^6$ yr the Sun was completely convective
($N^2<0$) and, therefore, no low-frequency g-modes
were excited, but only the high-frequency f- and
p-modes were excited. Subsequently, as the Sun
evolves (at times $t\approx1.5\times 10^6$ yr) a radiative region
(radiative core) where the Brunt - V$\ddot{\rm a}$is$\ddot {\rm a}$l$\ddot {\rm a}$ frequency
is greater than zero ($N^2>0$) appears and, consequently,
the excitation of low-frequency g-modes begins.
Only the structural part makes a major contribution
to the Brunt - V$\ddot{\rm a}$is$\ddot {\rm a}$l$\ddot {\rm a}$ frequency; the compositional
part is zero. This is because at such lifetimes
of the Sun hydrogen burning makes a minor
contribution to the energetics and evolution dynamics
of the Sun. In the course of subsequent evolution
the radiative region expands, the number of lowfrequency
g-modes increases, and their spectrum becomes
dense. The low-frequency g-modes begin to
play an increasingly important role. At times
$t\approx2\times10^7$ yr hydrogen burning begins to make an increasingly
large contribution to the total energy of the
star and, consequently, the compositional part of the
Brunt - V$\ddot{\rm a}$is$\ddot {\rm a}$l$\ddot {\rm a}$ frequency also begins to contribute to
the total frequency. In the course of subsequent evolution,
as hydrogen burns, this contribution becomes
progressively larger. One of the problems being investigated
here is to take into account the evolution of
the Brunt - V$\ddot{\rm a}$is$\ddot {\rm a}$l$\ddot {\rm a}$ frequency and the influence of this
evolution on the tidal interactions. The radiative region
expands approximately to a radius $r\approx0.73R_\odot$.
For the present Sun the boundary between the radiative
and convective regions is $r\approx0.712R_\odot$. In
Fig. 2 the Brunt - V$\ddot{\rm a}$is$\ddot {\rm a}$l$\ddot {\rm a}$ frequency is plotted against
radius for various stellar evolution times. It can be
seen from the plot that the Brunt - V$\ddot{\rm a}$is$\ddot {\rm a}$l$\ddot {\rm a}$ frequency
is subject to significant changes as the solar age increases;
both the frequency spectrum and its density
change accordingly, which can directly affect the tidal
forces. In Fig. 3 the Brunt - V$\ddot{\rm a}$is$\ddot {\rm a}$l$\ddot {\rm a}$ frequency is
plotted against radius for the present Sun. It follows
from this plot that the Brunt - V$\ddot{\rm a}$is$\ddot {\rm a}$l$\ddot {\rm a}$ frequency
exhibits a double-humped structure. The first and
second humps are associated with the compositional
and structural parts of the Brunt - V$\ddot{\rm a}$is$\ddot {\rm a}$l$\ddot {\rm a}$ frequency,
respectively. At a solar age $t\approx 8.61\times10^9$ yr the two
contributions become comparable.

The overlap integrals (1) are plotted against frequency
in the Cowling approximation in Fig. 4. For
a solar age $t\approx1.46\times 10^6$ yr, as has been said above,
only the f- and p-modes exist; therefore, there is no
low-frequency part of the spectrum. It can be seen
from the plot that the absolute value of the overlap
integrals decreases in the low-frequency part of the
spectrum as the Sun evolves. This decrease is gradual,
because the stellar structure does not change in
the evolution time of the Sun. On the main sequence
the Sun always has a radiative core and a convective
envelope. In the high-frequency part of the spectrum,
as the stellar age increases, the absolute value of
the overlap integrals increases and a characteristic
structure appears in the spectrumin the form of kinks.

\begin{figure}
\includegraphics[width=0.45\textwidth]{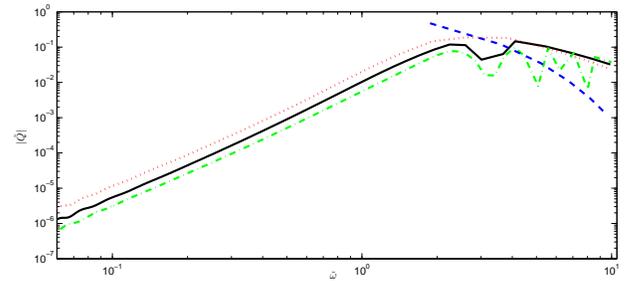}
\caption{Overlap integrals versus frequency for various solar ages: the dashed, dotted, solid, and dash-dotted lines
are for the ages $t=1.46\times 10^6$, $t=1.02\times 10^8$, $t=4.57\times 10^9$, and $t=8.61\times 10^9$ yr, respectively.}
\label{fig3}
\end{figure}

Figure 5 plots the change in the orbital period (in
days) of a planet with a mass of one Jupiter mass
around the Sun with time (in years). The solar evolution
models for which the overlap integrals were
calculated are presented in Table 1. At lifetimes of the
Sun $10^6<t<8\times10^6$ yr the spectrum at frequencies
$0.4<\tilde{\omega}<1$ is insufficiently dense. At other lifetimes
of the Sun the spectrum may be deemed sufficiently
dense with a good accuracy for all frequencies.
Therefore, Eq. (4) may be used. Equation (2) was
integrated over time from $t=3.50\times10^6$ to $t=8.61\times10^9$ yr.
The initial orbital periods of the planet around
the Sun were taken to be $P_{\rm orb}=2.6, 2.8, 3, 3.2$ days
(solid curves). It follows from Fig. 5 that in the
main-sequence lifetime of the Sun a planet with a
mass of one Jupiter mass with the initial orbital period
$P_{\rm orb}=2.8$ days will reduce its orbital period to $P_{\rm orb}=1.2$
days due to dynamical tides. The time scale of
the change in semimajor axis for this orbital period
$P_{\rm orb}=1.2$ days and this solar age $t=8.61\times10^9$ yr
is $T_a\approx1.5\times10^7$ yr. Consequently, it can be said
with confidence that this planet will fall onto the Sun
within several tens of millions of years. The same
analysis showed that a planet with the initial orbital
period $P_{\rm orb}=3$ days would reduce its orbital period
to $P_{\rm orb}\approx1.8$ days and would fall onto the Sun only
within approximately half a billion years.

For the initial orbital period $P_{\rm orb}=3$ days Fig. 5
compares the changes in orbital period with allowance
for the solar evolution (solid curve), without
allowance for the solar evolution (dash–dotted
curve), and the one calculated using Eq. (25) from
Essick and Weinberg (2016) (dotted curve). To calculate
the change in orbital period without allowance
for the solar evolution, we took the time scale at the
solar age $t=4.57\times10^9$ yr. All curves qualitatively
coincide, and the influence of solar evolution may
be disregarded in the estimation. This is because
hydrogen burning in the Sun is gradual, and no
change in structure occurs. Quantitatively, the dotted
curve from Essick and Weinberg (2016) gives smaller
changes in the planet’s orbital period. This may
be related to both the neglect of the change in the
time scales of the orbital parameters as the Sun
evolves and to a different approach in this paper to
the problem of the energy dissipation of tidally excited
modes associated with their nonlinear interactions.

\begin{figure}
\includegraphics[width=0.48\textwidth]{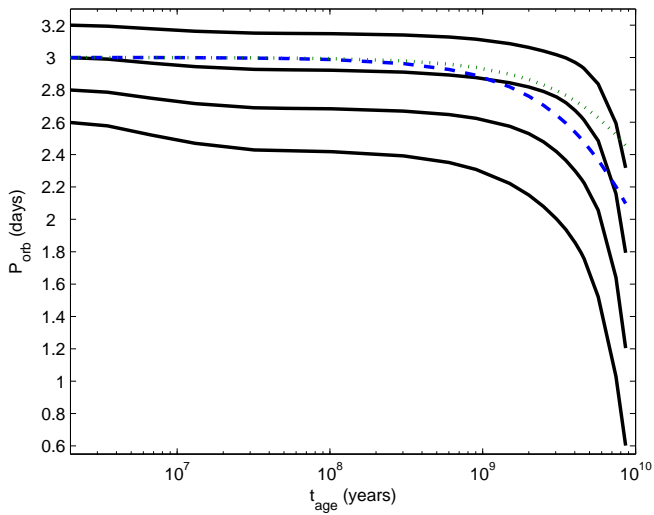}
\caption{Change in the orbital period of a planet with a mass of one Jupiter mass around the Sun due to dynamical
tides (for more details see the text).}
\label{fig35}
\end{figure}

\subsection{The Star with a Mass of 1.5 $M_\odot$}

In this subsection we present the results of our
calculations of the overlap integrals and the change
in orbital period for the star with a mass of one and
a half solar masses. The star is considered from the
time the protostar begins to collapse. This time is
quite arbitrary and is $t=10^{-5}$ yr; the initial radius is
$R=43.6 R_{\odot}$. The initial metallicity of the star is
$z=0.02$. At lifetimes of the star up to $t\approx10^6$ yr it is completely
convective ($N^2<0$). In such a star no lowfrequency
g-modes are excited, and only the highfrequency
f- and p-modes are present; therefore, the
effective energy and angular momentum transfer from
the orbit to the star will be suppressed. As the stellar
age increases further, hydrogen burning begins at the
stellar center and a radiative core appears ($N^2>0$),
which begins to expand, encompassing progressively
newer layers. The star has a well-defined convective
envelope and a radiative core. The radiative region
expands up to $R\approx0.8R_\odot$ for the age $t=10^7$ yr.
A fairly dense spectrum of low-frequency g-modes
appears very rapidly in such a star.

At times $t\approx2\times 10^7$ yr the stellar structure
changes completely. A convective core appears
in the star, while its envelope becomes completely
radiative. The star abruptly contracts from a radius
$R\approx2.02R_\odot$ to $R\approx1.5R_\odot$. The evolution time scales
of the orbital parameters (2) increase sharply. The star
passes from a solar-type star that has a radiative core
and a convective envelope to a star with a convective
core and a radiative envelope.

\begin{figure}
\includegraphics[width=0.45\textwidth]{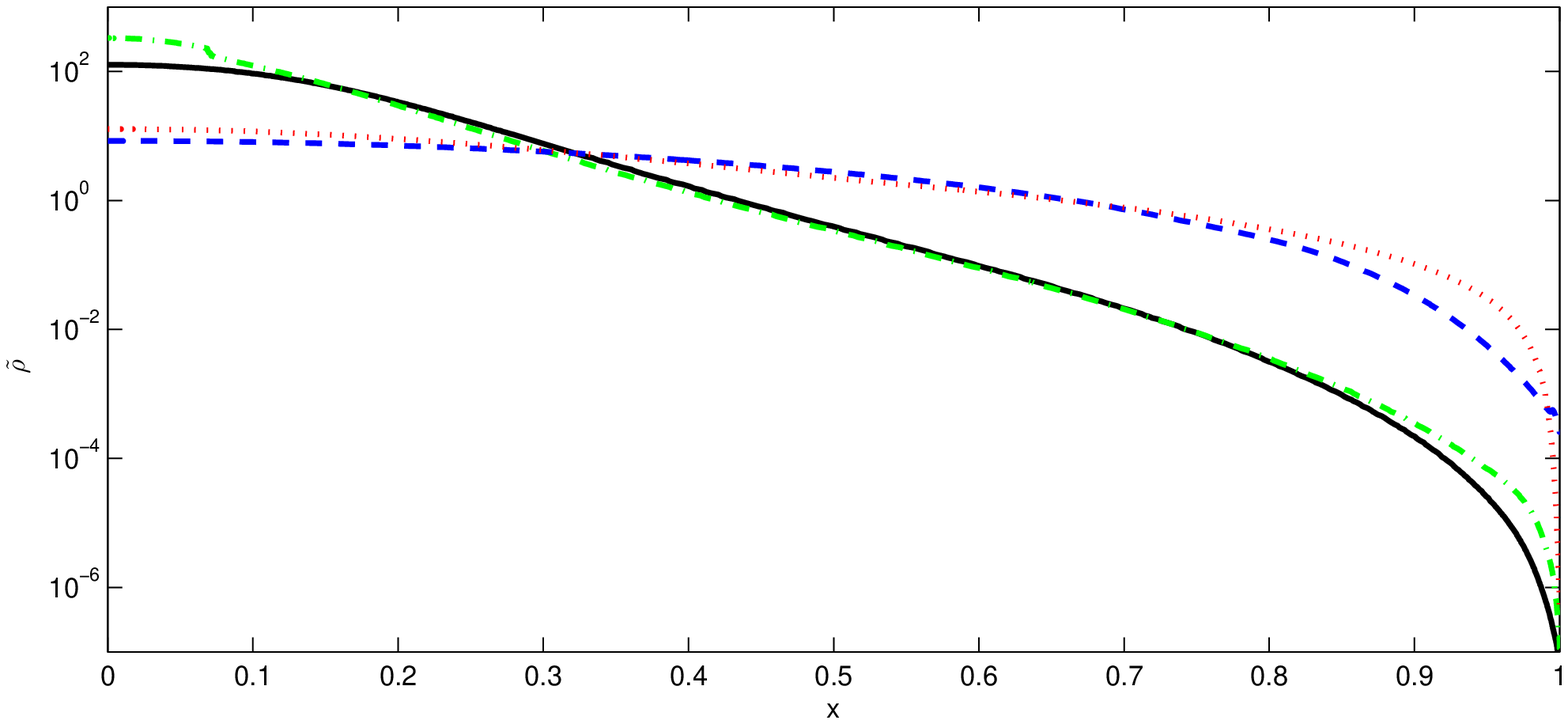}
\includegraphics[width=0.45\textwidth]{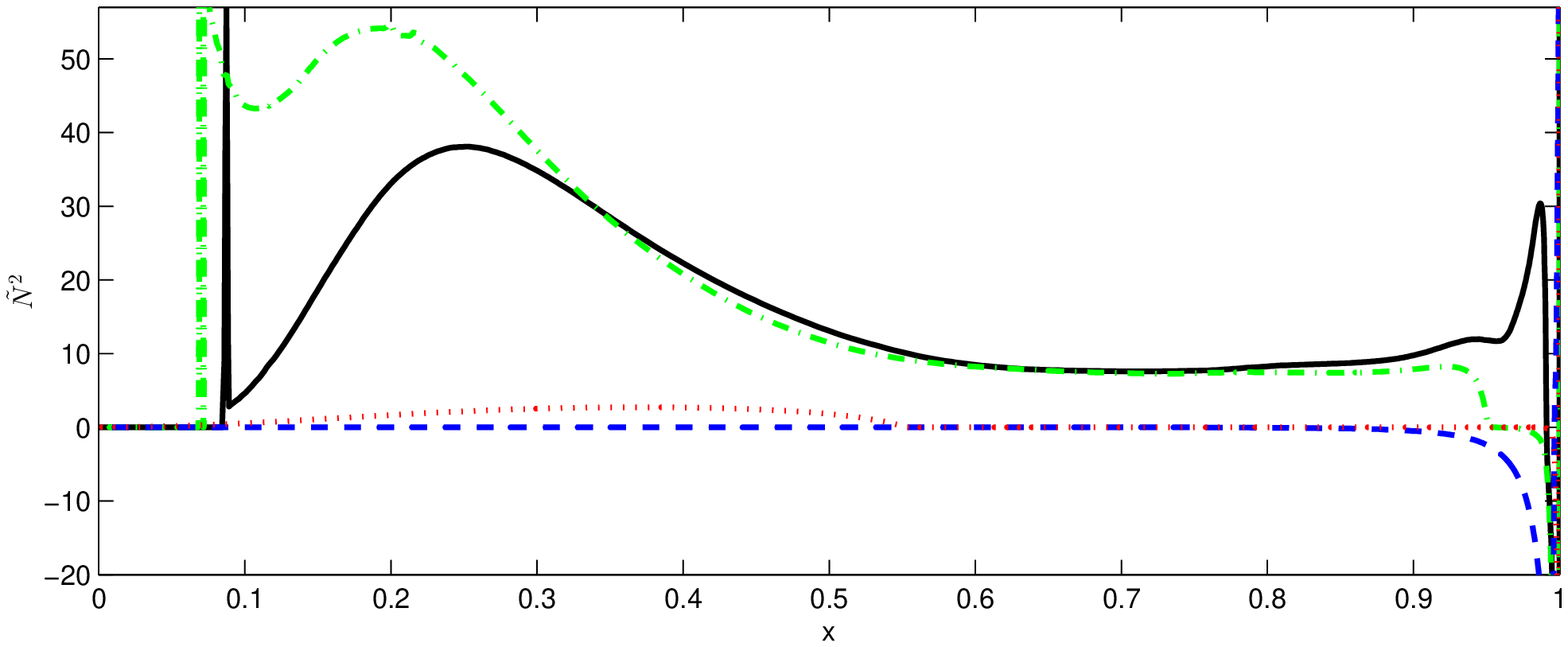}
\caption{Density (a) and Brunt - V$\ddot{\rm a}$is$\ddot {\rm a}$l$\ddot {\rm a}$ frequency (b) for the star of one and a half solar masses versus radius for
various times: the dashed, dotted, solid, and dash–dotted lines are for the ages $t=1.08\times 10^2$, $t=4.60\times 10^6$, $t=1.36\times 10^8$, and
$t=1.47\times 10^9$ yr, respectively.}
\label{fig4}
\end{figure}

In Fig. 6 the stellar density and Brunt - V$\ddot{\rm a}$is$\ddot {\rm a}$l$\ddot {\rm a}$
frequency are plotted against radius for various times.
From the figure we see how the central density increased
as the star contracted. It can be seen from the
figure for the Brunt - V$\ddot{\rm a}$is$\ddot {\rm a}$l$\ddot {\rm a}$ frequency that the star
was initially completely convective (dashed curve)
and subsequently a solar-type one (the envelope was convective, and the core was radiative; the dotted curve). At the subsequent times the star has a more
complex structure, the core becomes convective, and
the envelope is completely radiative (solid curve). A
thin convective envelope then again appears in its
envelope (dash–dotted curve).

\begin{figure}
\includegraphics[width=0.45\textwidth]{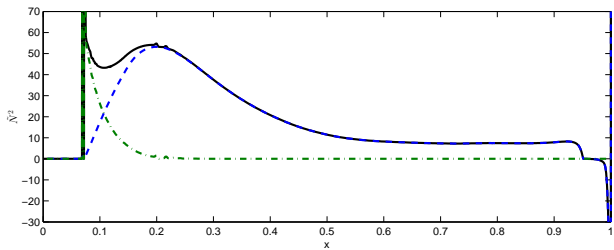}
\caption{Brunt - V$\ddot{\rm a}$is$\ddot {\rm a}$l$\ddot {\rm a}$ frequency versus radius for the star of one and a half solar masses for the age $t=1.47\times 10^9$ yr: the solid, dashed, and dash-dotted lines indicate the total Brunt - V$\ddot{\rm a}$is$\ddot {\rm a}$l$\ddot {\rm a}$ frequency $N^2$, the structural part $N^2_{\rm st}$, and the compositional part $N^2_{\rm com}$, respectively.}
\label{fig5}
\end{figure}

Figure 7 shows the Brunt - V$\ddot{\rm a}$is$\ddot {\rm a}$l$\ddot {\rm a}$ frequency as
well as its structural and compositional parts. It
follows from the figure that the compositional part
makes the greatest contribution at the boundary between
the convective and radiative regions.

\begin{figure}
\includegraphics[width=0.48\textwidth]{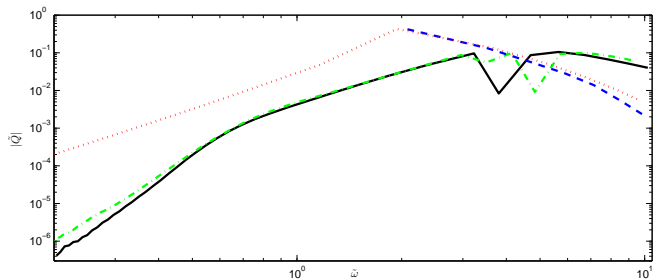}
\caption{Overlap integrals versus frequency for the star of one and a half solar masses: the dashed, dotted, solid,
and dash–dotted lines are for the ages $t=1.08\times 10^2$, $t=4.60\times 10^6$, $t=1.36\times 10^8$, and $t=1.47\times 10^9$ yr, respectively.}
\label{fig6}
\end{figure}

The overlap integrals are plotted against frequency
in Fig. 8. The dashed line does not extend to the lowfrequency
part of the spectrum, because the star at
this time is completely convective ($N^2<0$). It can
also be seen from the figure that the absolute value of
the overlap integrals rapidly drops by more than two
orders of magnitude as the star’s lifetime increases.
Such catastrophic changes are due to the change
in stellar structure. In the high-frequency part the
absolute value of the overlap integrals increases by an
order of magnitude, and characteristic kinks appear
in the spectrum.

Figure 9 plots the change in the orbital period (in
days) of a planet with a mass of one Jupiter mass
around the star of one and a half solar masses with
time (in years). The stellar evolutionmodels for which
the overlap integrals were calculated are presented in
Table 2. Equation (2) was integrated over time from
$t\approx1.98\times10^7$ to $t\approx2.24\times10^9$ yr. For such a stellar
lifetime interval the g-mode spectrum is sufficiently
dense. The initial orbital periods of the planet around
the Sun are $P_{\rm orb}=1.2, 1.3, 1.4, 2, 2.5, 3$ days (solid
curves). It follows from the figure that all planets
with a mass equal to one Jupiter mass and with an
initial orbital period of 2 days or shorter will fall onto
the star in the main-sequence lifetime of the star. A
planet with the initial orbital period $P_{\rm orb}=2$ days will
fall onto the star, while a planet with the initial orbital
period $P_{\rm orb}=2.5$ days will reduce its orbital period to
$P_{\rm orb}=1.9$ days in the main-sequence lifetime of the
star. The dashed curve indicates the change in orbital
period without allowance for the stellar evolution. The
time scale of the orbital change was taken at the star’s
lifetime $t=7.82\times10^8$ yr. It can be seen from Fig. 9
that the planet does not fall onto the star in the mainsequence
lifetime of the star without allowance for its
evolution.

\begin{figure}
\includegraphics[width=0.48\textwidth]{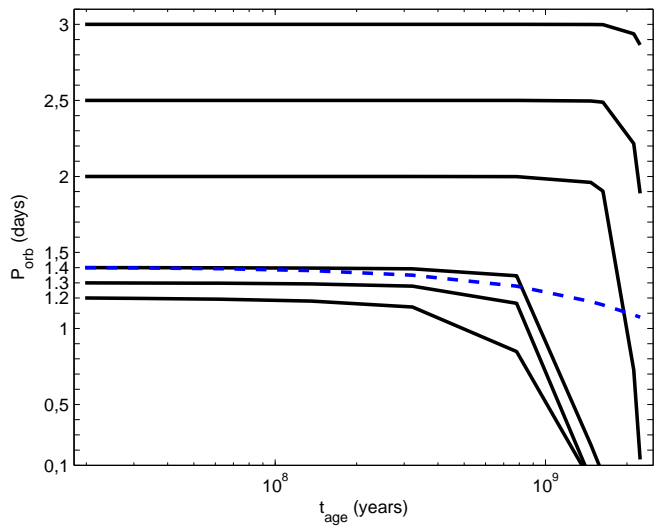}
\caption{Change in the orbital period of a planet with a mass of one Jupiter mass around the star of one and a half solar masses due to dynamical tides (for more details see the text).}
\label{fig65}
\end{figure}

\subsection{The Star with a Mass of 2$M_\odot$}

The last star that we will consider is the star with
a mass of two solar masses. The time the protostar
begins to collapse is $t\approx10^{-5}$ yr, the initial stellar
radius is $R=57.4 R_\odot$, and the stellar metallicity is
$z=0.02$. At initial times up to $t\approx3\times10^5$ yr the star
is completely convective ($N^2<0$). No low-frequency
g-modes are excited; only the high-frequency f- and
p-modes are present. The dynamical tides play no
significant role in such a star. However, as the star
evolves further, at times $t\approx3\times10^5$ yr a radiative
region (radiative core) appears in the star, which begins
to expand. The low-frequency g-modes are well
excited in this region ($N^2>0$), and the dynamical
tides begin to play an increasingly significant role as
the star evolves further. Such a star resembles solartype
stars that have a radiative core and a convective
envelope. The radiative region expands quite rapidly,
and this region almost reaches the stellar surface
$R=0.987R_\odot$ already at times $t\approx 5.7\times10^6$ yr. The
star becomes almost completely radiative; the Brunt - V$\ddot{\rm a}$is$\ddot {\rm a}$l$\ddot {\rm a}$ frequency in the entire region becomes positive
($N^2>0$) in the entire region, except for the stellar
surface. A sufficiently dense g-mode spectrum is
generated in such a star. In Fig. 10 the stellar density
and Brunt - V$\ddot{\rm a}$is$\ddot {\rm a}$l$\ddot {\rm a}$ frequency are plotted against
radius for various times.

\begin{figure}
\includegraphics[width=0.45\textwidth]{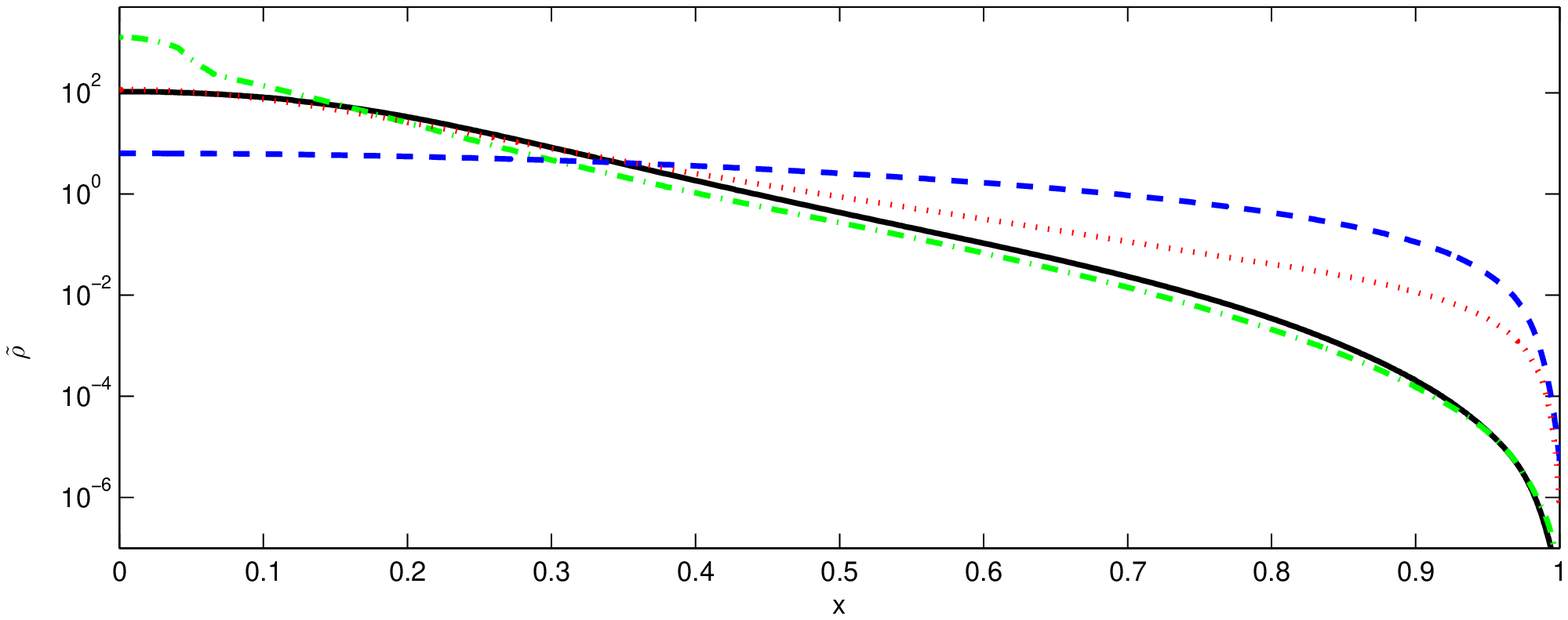}
\includegraphics[width=0.45\textwidth]{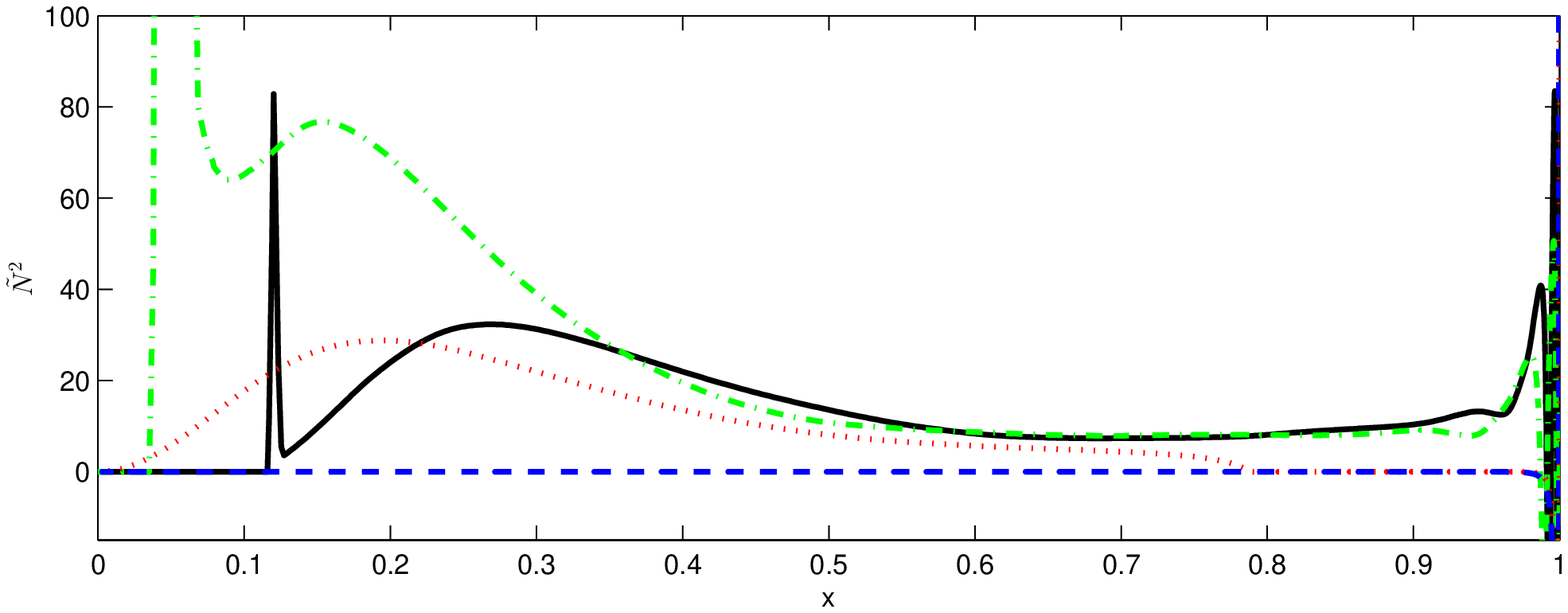}
\caption{Density (a) and Brunt - V$\ddot{\rm a}$is$\ddot {\rm a}$l$\ddot {\rm a}$ frequency (b) for the star of two solar masses versus radius for
various times: the dashed, dotted, solid, and dash–dotted lines are for the ages $t=1.49\times 10^5$, $t=4.54\times 10^6$, $t=9.50\times 10^7$, and
$t=9.09\times 10^8$, respectively.}
\label{fig7}
\end{figure}

As the lifetime increases further ($t>5.8\times10^6$ yr),
a convective core ($N^2<0$) appears in the star,
which begins to expand as the star evolves. The
compositional part of the Brunt - V$\ddot{\rm a}$is$\ddot {\rm a}$l$\ddot {\rm a}$ frequency
$N^2_{\rm com}$ begins to play an increasingly large role at
the boundary between the convective and radiative
regions (Fig. 11).

The stellar structure does not undergo significant
changes as the lifetime increases further. Only the
convective core evolves. Initially, the convective region
expands, and then its expansion is replaced by
contraction (Fig. 10).

\begin{figure}
\includegraphics[width=0.45\textwidth]{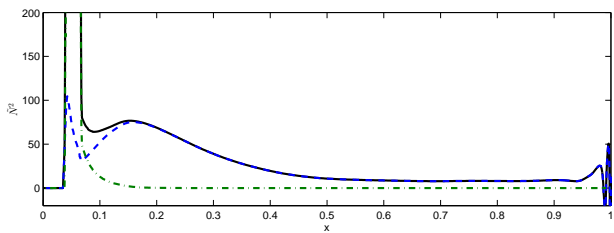}
\caption{Brunt - V$\ddot{\rm a}$is$\ddot {\rm a}$l$\ddot {\rm a}$ frequency versus radius for the star of two solar masses for the age $t=9.09\times 10^8$ yr:
the solid, dashed, and dash-dotted lines indicate the total Brunt - V$\ddot{\rm a}$is$\ddot {\rm a}$l$\ddot {\rm a}$ frequency $N^2$, the structural part $N^2_{\rm st}$, and the
compositional part $N^2_{\rm com}$, respectively.}
\label{fig8}
\end{figure}

\begin{figure}
\includegraphics[width=0.45\textwidth]{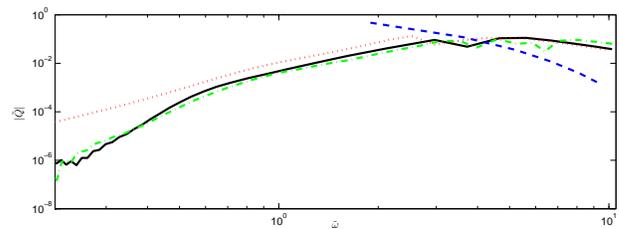}
\caption{Overlap integrals versus frequency for the star of two solar masses: the dashed, dotted, solid, and
dash-dotted lines are for the ages $t=1.49\times 10^5$, $t=4.54\times 10^6$, $t=9.50\times 10^7$, and $t=9.09\times 10^8$ yr, respectively.}
\label{fig9}
\end{figure}

The overlap integrals are plotted against frequency
in Fig. 12. The dashed line does not extend to the
low-frequency part of the spectrum, because the star
at this time is completely convective ($N^2<0$). As
the star’s lifetime increases, the absolute value of the
overlap integral rapidly drops by more than an order
of magnitude. Such catastrophic changes are due
to the change in stellar structure. When the stellar
structure undergoes no significant changes, the overlap
integrals change gradually and insignificantly (see
Fig. 12, the solid and dash-dotted lines).

Figure 13 plots the change in the orbital period (in
days) of a planet with a mass of one Jupiter mass
around the star of two solar masses with time (in
years). The evolution models of the star of two solar
masses for which the overlap integrals were calculated
are presented in Table 3. Equation (2) was
integrated over time from $t\approx6.29\times10^6$ to $t\approx9.13\times10^8$
yr. For such a stellar lifetime interval the g-mode
spectrum is sufficiently dense. The initial orbital
periods of the planet around the Sun were taken to
be $P_{\rm orb}=1.5, 1.6, 1.7, 2, 2.5, 3$ days (solid curves). It
follows from the figure that all planets with a mass
equal to one Jupiter mass and with an initial orbital
period of 2 days or shorter will fall onto the star in
the main-sequence lifetime of the star (Fig. 13). A
planet with the initial orbital period $P_{\rm orb}=2.5$ days
will reduce its orbital period only to $P_{\rm orb}\approx2.35$ days
in the main-sequence lifetime of the star. The dashed
curve indicates the change in orbital period without
allowance for the stellar evolution. The time scale
of the orbital change was taken at the star's lifetime
$t=3.01\times10^8$ yr. As can be seen from Fig. 13, the
planet does not fall onto the star in the main-sequence
lifetime of the star without allowance for its evolution. q

\begin{figure}
\includegraphics[width=0.48\textwidth]{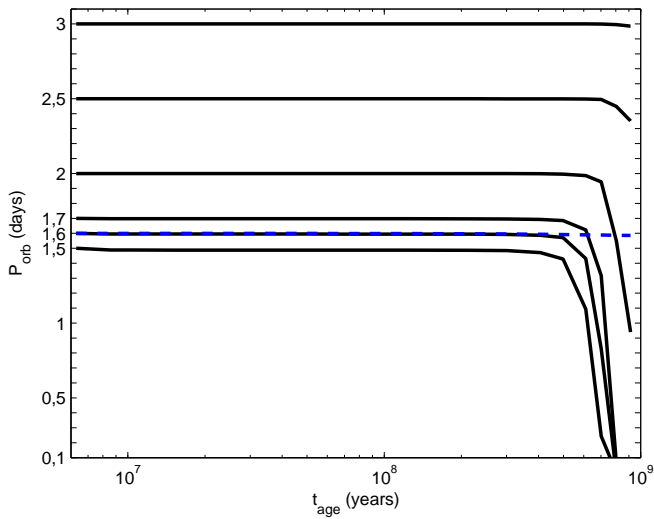}
\caption{Change in the orbital period of a planet with a mass of one Jupiter mass around the star of two solar
masses due to dynamical tides (for more details see the text).}
\label{fig95}
\end{figure}

Just as for the star of one and a half solar masses,
the evolution time scales (3) of the planet decrease
as the star begins to move off the main sequence.
This is because the hydrogen reserves at the stellar
center begin to be depleted, and the convective core
becomes a helium one. This leads to an expansion
of the star and, consequently, to an increase in the
number of g-modes and the density of the spectrum,
which leads to an increase in the pumping of orbital
energy into the energy of stellar oscillations and to the
fall of short-period planets onto the star.

\section{CONCLUSIONS}

We considered the change in the orbital period of
a planet around a star with a mass of one, one and
a half, and two solar masses. The behavior of the
overlap integrals for these models as a function of
the main-sequence lifetime of the star was analyzed.
We showed that for the same star with a different
structure during its evolution the overlap integrals
could differ by two or more orders of magnitude.

We showed that all planets with a mass of one
Jupiter mass that revolve around the Sun with an
orbital period $P_{\rm orb}\approx2.8$ days or shorter should fall
onto the stellar surface in the main-sequence lifetime
of the star. All planets with a mass of one Jupiter
mass that revolve around the stars with a mass of one
and a half and two solar masses with an orbital period
$P_{\rm orb}\approx2$ or shorter will fall onto the stellar surface in
the main-sequence lifetime of the star. Such a fall
may not occur without allowance for the evolution of
the star itself.

These results qualitatively agree with the previous
results from Bolmont and Mathis (2016) and
Penev (2012). Calculations for the stellar models
considered in this paper should be made for a more
accurate quantitative estimate.

\section{ACKNOWLEDGMENTS}

I am grateful to P.B. Ivanov who read the paper for
a number of valuable remarks and to J. Papaloizou for
a fruitful discussion. This work was financially supported
by the Russian Foundation for Basic Research
(project nos. 140200831, 150208476, 160201043),
grant no. NSh-6595.2016.2 from the President of
the Russian Federation for State Support of Leading
Scientific Schools, and Program 7 of the Presidium of
the Russian Academy of Sciences.

\section{APPENDIX A}

The system of equations describing the adiabatic
perturbations in stars is written in dimensionless form
(Christensen-Dalsgaard 1998):
\begin{eqnarray}
 \frac{dy_1}{dx} &=& -\left(\frac{2}{x}-\frac{\tilde{g}_0}{\tilde{c}^2_0}\right)y_1+\nonumber\\
 &+& \left(\frac{1}{x}-\frac{\tilde{\omega}^2 x}{\tilde{c}^2_0l(l+1)}\right)
 y_2+\frac{\tilde{g}_0}{\tilde{c}^2_0}y_3,\nonumber\\
 \frac{dy_2}{dx} &=& \left(1-\frac{\tilde{N}^2_0}{\tilde{\omega}^2}\right)\frac{l(l+1)}{x}
 y_1-\frac{y_2}{x}+\frac{\tilde{N}^2_0}{\tilde{g}_0}y_2-\nonumber\\
 &-& \frac{l(l+1)\tilde{N}^2_0}{\tilde{\omega}^2 x}y_3,\nonumber\\
 \frac{dy_3}{dx} &=& \frac{y_3}{x}+\frac{y_4}{x},\nonumber\\
 \frac{dy_4}{dx} &=& \frac{3x\tilde{\rho}_0 \tilde{N}^2_0}{\tilde{g}_0^2}y_1+
 \frac{3x^2\tilde{\omega}^2\tilde{\rho}_0}{\tilde{g}_0\tilde{c}^2_0l(l+1)}y_2\nonumber\\
 &+& \left(\frac{l(l+1)}{x}-\frac{6\tilde{\rho}_0}{\tilde{g}_0}
 +
 \frac{3x\tilde{N}^2_0\tilde{\rho}_0}{\tilde{g}_0^2}\right)y_3-\nonumber\\
 &-& 2\left(-\frac{1}{x}+\frac{3\tilde{\rho}_0}{\tilde{g}_0}\right)y_4,
 \label{dif}
\end{eqnarray}
where the following dimensionless variables are introduced:
$x=\frac{r}{R}$ is the radius, $y_1=\frac{\xi_r}{R}$ is the
radial perturbation, $y_2=\left(\frac{p^{'}}{\rho_0}+\Phi^{'}\right)\frac{l(l+1)}{\omega^2 r^2}x=l(l+1)\frac{\xi_h}{R}$ is the horizontal perturbation, $y_3=x\frac{\Phi^{'}}{g_0 r}$ and
$y_4=x^2\frac{d}{dx}\left(\frac{y_3}{x}\right)$ are the perturbations of the gravitational
potential and its derivative, respectively. The
subscript $0$ refers to an unperturbed quantity, and the
prime $'$ refers to a perturbed quantity. A tilde over a
quantity means that this quantity is dimensionless.
The normalization is done as follows:
\begin{eqnarray}
 \omega^2 &=& \frac{GM}{R^3}\tilde{\omega}^2, \quad c^2_0=\frac{GM}{R}\tilde{c}^2_0,\quad
 g_0=\frac{GM}{R^2}\tilde{g}_0,\nonumber\\
 N^2_0 &=& \frac{GM}{R^3}\tilde{N}^2_0,\qquad
 \rho_0 = \tilde{\rho_0}\frac{3M}{4\pi R^3}, \nonumber\\
 p_0 &=& \frac{GM^2}{4\pi R^4}\tilde{p}_0,\qquad
 \tilde{c}^2_0=\frac{\Gamma \tilde{p}_0}{3\tilde{\rho}_0},
\end{eqnarray}
where $c^2$ is the sound speed squared, $p$ is the pressure,
$\rho$ is the density, and $N^2$ is the Brunt - V$\ddot{\rm a}$is$\ddot {\rm a}$l$\ddot {\rm a}$
frequency squared. Four boundary conditions should
be added to the four first-order differential equations.
Two boundary conditions are specified at the stellar
center, and the other two are specified on the stellar
surface (Christensen-Dalsgaard 1998). The boundary
conditions at the stellar center $x\rightarrow0$ are
\begin{eqnarray}
 (l+1)y_1=y_2,\quad y_4=(l-2)y_3.
\end{eqnarray}
The boundary conditions on the stellar surface
$x\rightarrow1$  are
\begin{eqnarray}
 y_2=\frac{l(l+1)}{\tilde{\omega}^2}(y_1+y_3),\qquad
 ly_3=-y_4.
\end{eqnarray}
There exists the Cowling (1941) approximation, in
which the perturbed gravitational potential is neglected,
to investigate the low-frequency g-modes.
This approximation holds good and simplifies considerably
the system of equations for the adiabatic
perturbations (14). In the Cowling approximation the
system of equations (14) will be rewritten in a simpler
form:
\begin{eqnarray}
 \frac{dy_1}{dx} &=& -\left(\frac{2}{x}-\frac{\tilde{g}_0}{\tilde{c}^2_0}\right)y_1+
 \left(\frac{1}{x}-\frac{\tilde{\omega}^2 x}{\tilde{c}^2_0l(l+1)}\right)
 y_2,\nonumber\\
 \frac{dy_2}{dx} &=& \left(1-\frac{\tilde{N}^2_0}{\tilde{\omega}^2}\right)\frac{l(l+1)}{x}
 y_1-\frac{y_2}{x}+\frac{\tilde{N}^2_0}{\tilde{g}_0}y_2.
 \label{dif1}
\end{eqnarray}
In the Cowling approximation the system of equations
(18) ceases to depend explicitly on the stellar
density. Thus, in the low-frequency limit the radial
and horizontal displacements do not depend on the
stellar density. In contrast, the overlap integrals (1)
directly depend on the density. In the Cowling approximation
not four but two boundary conditions
should be added to the system of equations (18). One
boundary condition is specified at the stellar center,
and the other one is specified on the stellar surface.
The boundary condition at the stellar center $x\rightarrow0$ is
\begin{eqnarray}
 (l+1)y_1=y_2.
\end{eqnarray}
The boundary condition on the stellar surface $x\rightarrow 1$
is
\begin{eqnarray}
 y_2=\frac{l(l+1)}{\tilde{\omega}^2}y_1.
\end{eqnarray}
To determine the eigenfrequency spectrum of the system
of equations (14) or (18), this system should be
redefined. This requires that the determinant composed
of the eigenfunctions of the system of equations
(14) or (18) be equal to zero. This will hold
only for some frequencies (eigenfrequencies) of the
system of equations (14) or (18), which determines
the perturbation spectrum.

\section{APPENDIX B}

The tables present the models for which the overlap
integrals (1) and the evolution time scales of the
orbital parameters (2) were calculated. The notation
is as follows: $t$ is the time in years, $R$ is the radius
measured in solar radii, $L$ is the luminosity measured
in solar luminosities, $T_{eff}$ is effective surface temperature
measured in Kelvins, h1 is the hydrogen mass
fraction at the stellar center, and $\rho_m=\frac{3M}{4\pi R^3}$
is the mean stellar density measured in $g/cm^3$.

\begin{table}
\caption{The models used to calculate the characteristic
orbital parameters for the Sun}
\label{tabSun}
\begin{ruledtabular}
\begin{tabular}{|c|c|c|c|c|c|}
 t & R/$R_\odot$ & L/$L_\odot$ & $T_{eff}\times10^3$ & $h_1$ & $\rho_m$\\
\hline
 1.46$\times10^6$ & 2.02 & 1.42 & 4.43 & 0.7 & 0.17\\
\hline
 1.88$\times10^6$ & 1.88 & 1.20 & 4.42 & 0.7 & 0.21\\
\hline
 2.38$\times10^6$ & 1.75 & 1.02 & 4.39 & 0.7 & 0.26\\
\hline
 3.50$\times10^6$ & 1.56 & 0.79 & 4.35 & 0.7 & 0.37\\
\hline
 6.79$\times10^6$ & 1.30 & 0.53 & 4.31 & 0.7 & 0.64\\
\hline
 1.31$\times10^7$ & 1.12 & 0.44 & 4.45 & 0.7 & 1.00\\
\hline
 3.19$\times10^7$ & 0.99 & 0.90 & 5.65 & 0.7 & 1.45\\
\hline
 5.26$\times10^7$ & 0.88 & 0.69 & 5.62 & 0.7 & 2.07\\
\hline
 1.03$\times10^8$ & 0.88 & 0.70 & 5.63 & 0.7 & 2.07\\
\hline
 3.03$\times10^8$ & 0.89 & 0.72 & 5.64 & 0.68 & 2.00\\
\hline
 6.03$\times10^8$ & 0.89 & 0.73 & 5.65 & 0.66 & 2.00\\
\hline
 9.03$\times10^8$ & 0.90 & 0.75 & 5.66 & 0.64 & 1.94\\
\hline
 1.50$\times10^9$ & 0.91 & 0.78 & 5.68 & 0.59 & 1.87\\
\hline
 2.00$\times10^9$ & 0.92 & 0.81 & 5.70 & 0.55 & 1.81\\
\hline
 2.50$\times10^9$ & 0.94 & 0.84 & 5.71 & 0.51 & 1.70\\
\hline
 3.04$\times10^9$ & 0.95 & 0.87 & 5.73 & 0.47 & 1.65\\
\hline
 3.50$\times10^9$ & 0.96 & 0.91 & 5.74 & 0.43 & 1.60\\
\hline
 4.03$\times10^9$ & 0.98 & 0.95 & 5.75 & 0.39 & 1.50\\
\hline
 4.40$\times10^9$ & 0.99 & 0.98 & 5.76 & 0.36 & 1.45\\
\hline
 4.50$\times10^9$ & 1.00 & 0.98 & 5.76 & 0.35 & 1.41\\
\hline
 4.57$\times10^9$ & 1.00 & 0.99 & 5.76 & 0.34 & 1.41\\
\hline
 4.61$\times10^9$ & 1.00 & 0.99 & 5.77 & 0.34 & 1.41\\
\hline
 5.70$\times10^9$ & 1.04 & 1.09 & 5.80 & 0.24 & 1.25\\
\hline
 7.44$\times10^9$ & 1.13 & 1.30 & 5.81 & 0.08 & 0.98\\
\hline
 8.61$\times10^9$ & 1.21 & 1.48 & 5.79 & 0.002 & 0.80
\end{tabular}
\end{ruledtabular}
\end{table}

\begin{table}
\caption{The models used to calculate the characteristic
orbital parameters for the star of one and a half solar
masses}
\label{tab1.5}
\begin{ruledtabular}
\begin{tabular}{|c|c|c|c|c|c|}
 t & R/$R_\odot$ & L/$L_\odot$ & $T_{eff}\times10^3$ & $h_1$ & $\rho_m$\\
\hline
 1.00$\times10^{-5}$ & 43.6 & 782 & 4.62 & 0.7 & 2.55$\times10^{-5}$\\
\hline
 5.37$\times10^{-5}$ & 43.6 & 765 & 4.60 & 0.7 & 2.55$\times10^{-5}$\\
\hline
 1.49$\times10^{-4}$ & 43.6 & 737 & 4.56 & 0.7 & 2.55$\times10^{-5}$\\
\hline
 4.85$\times10^{-4}$ & 43.5 & 650 & 4.42 & 0.7 & 2.57$\times10^{-5}$\\
\hline
 1.06$\times10^{-3}$ & 43.5 & 574 & 4.49 & 0.7 & 2.57$\times10^{-5}$\\
\hline
 5.67$\times10^{-3}$ & 43.3 & 417 & 3.97 & 0.7 & 2.61$\times10^{-5}$\\
\hline
 1.18$\times10^{-2}$ & 43.3 & 394 & 3.91 & 0.7 & 2.61$\times10^{-5}$\\
\hline
 5.10$\times10^{-2}$ & 43.2 & 386 & 3.89 & 0.7 & 2.63$\times10^{-5}$\\
\hline
 1.06$\times10^{-1}$ & 43.1 & 377 & 3.88 & 0.7 & 2.64$\times10^{-5}$\\
\hline
 5.46$\times10^{-1}$ & 43.0 & 366 & 3.86 & 0.7 & 2.66$\times10^{-5}$\\
\hline
 1.13 & 42.8 & 362 & 3.85 & 0.7 & 2.70$\times10^{-5}$\\
\hline
 5.84 & 42.6 & 358 & 3.85 & 0.7 & 2.74$\times10^{-5}$\\
\hline
 1.01$\times10$ & 42.5 & 357 & 3.85 & 0.7 & 2.76$\times10^{-5}$\\
\hline
 5.21$\times10$ & 41.7 & 346 & 3.86 & 0.7 & 2.92$\times10^{-5}$\\
\hline
 1.08$\times10^2$ & 40.6 & 333 & 3.87 & 0.7 & 3.16$\times10^{-5}$\\
\hline
 5.56$\times10^2$ & 34.7 & 261 & 3.94 & 0.7 & 5.07$\times10^{-5}$\\
\hline
 1.09$\times10^3$ & 30.5 & 215 & 4.01 & 0.7 & 7.46$\times10^{-5}$\\
\hline
 4.64$\times10^3$ & 20.1 & 113 & 4.20 & 0.7 & 2.61$\times10^{-4}$\\
\hline
 1.39$\times10^4$ & 13.6 & 61.1 & 4.37 & 0.7 & 8.42$\times10^{-4}$\\
\hline
 5.31$\times10^4$ & 8.26 & 26.8 & 4.57 & 0.7 & 3.8$\times10^{-3}$\\
\hline
 1.06$\times10^5$ & 6.37 & 17.2 & 4.66 & 0.7 & 8.2$\times10^{-3}$\\
\hline
 4.28$\times10^5$ & 3.82 & 6.90 & 4.79 & 0.7 & 3.8$\times10^{-2}$\\
\hline
 8.81$\times10^5$ & 2.96 & 4.22 & 4.81 & 0.7 & 8.2$\times10^{-2}$\\
\hline
 1.31$\times10^6$ & 2.58 & 3.23 & 4.82 & 0.7 & 0.12\\
\hline
 3.17$\times10^6$ & 1.97 & 1.95 & 4.86 & 0.7 & 0.28\\
\hline
 4.60$\times10^6$ & 1.80 & 1.77 & 4.97 & 0.7 & 0.36\\
\hline
 7.38$\times10^6$ & 1.70 & 2.10 & 5.32 & 0.7 & 0.43\\
\hline
 8.62$\times10^6$ & 1.75 & 2.61 & 5.54 & 0.7 & 0.40\\
\hline
 1.06$\times10^7$ & 2.02 & 4.67 & 5.98 & 0.7 & 0.26\\
\hline
 1.98$\times10^7$ & 1.53 & 4.90 & 6.96 & 0.7 & 0.59\\
\hline
 3.18$\times10^7$ & 1.47 & 4.81 & 7.05 & 0.7 & 0.67\\
\hline
 5.34$\times10^7$ & 1.47 & 4.82 & 7.06 & 0.69 & 0.67\\
\hline
 6.17$\times10^7$ & 1.47 & 4.82 & 7.06 & 0.69 & 0.67\\
\hline
 1.36$\times10^8$ & 1.48 & 4.88 & 7.06 & 0.67 & 0.65\\
\hline
 3.22$\times10^8$ & 1.51 & 5.04 & 7.04 & 0.63 & 0.61\\
\hline
 7.82$\times10^8$ & 1.61 & 5.42 & 6.94 & 0.53 & 0.51\\
\hline
 1.47$\times10^9$ & 1.86 & 5.91 & 6.61 & 0.30 & 0.33\\
\hline
 1.63$\times10^9$ & 1.93 & 5.99 & 6.51 & 0.24 & 0.29\\
\hline
 2.12$\times10^9$ & 2.16 & 6.98 & 6.38 & 6.8$\times10^{-3}$ & 0.21\\
\hline
 2.24$\times10^9$ & 2.46 & 8.79 & 6.35 & 5.6$\times10^{-11}$ & 0.14
\end{tabular}
\end{ruledtabular}
\end{table}

\begin{table}
\caption{The models used to calculate the characteristic
orbital parameters for the star of two solar masses}
\label{tab2}
\begin{ruledtabular}
\begin{tabular}{|c|c|c|c|c|c|}
 t & R/$R_\odot$ & L/$L_\odot$ & $T_{eff}\times10^3$ & $h_1$ & $\rho_m$\\
\hline
 $10^{-5}$ & 57.4 & 1170 & 4.46 & 0.7 & 1.49$\times10^{-5}$\\
\hline
 4.55$\times10^{-1}$ & 56.4 & 633 & 3.86 & 0.7 & 1.57$\times10^{-5}$\\
\hline
 3.35$\times10^2$ & 47.3 & 483 & 3.94 & 0.7 & 2.67$\times10^{-5}$\\
\hline
 1.37$\times10^3$ & 35.3 & 311 & 4.08 & 0.7 & 6.42$\times10^{-5}$\\
\hline
 7.83$\times10^3$ & 19.8 & 128 & 4.37 & 0.7 & 3.64$\times10^{-5}$\\
\hline
 3.49$\times10^4$ & 11.3 & 52.5 & 4.63 & 0.7 & 2.0$\times10^{-3}$\\
\hline
 1.49$\times10^5$ & 6.51 & 21.0 & 4.85 & 0.7 & 1.02$\times10^{-2}$\\
\hline
 3.07$\times10^5$ & 4.96 & 13.2 & 4.94 & 0.7 & 2.31$\times10^{-2}$\\
\hline
 6.37$\times10^5$ & 3.81 & 8.18 & 5.01 & 0.7 & 5.10$\times10^{-2}$\\
\hline
 1.36$\times10^6$ & 2.97 & 5.30 & 5.09 & 0.7 & 0.108\\
\hline
 3.13$\times10^6$ & 2.51 & 4.82 & 5.40 & 0.7 & 0.179\\
\hline
 4.54$\times10^6$ & 2.99 & 9.49 & 5.86 & 0.7 & 0.106\\
\hline
 5.21$\times10^6$ & 3.26 & 14.6 & 6.25 & 0.7 & 8.15$\times10^{-2}$\\
\hline
 5.68$\times10^6$ & 2.99 & 19.5 & 7.02 & 0.7 & 0.106\\
\hline
 6.29$\times10^6$ & 2.36 & 23.8 & 8.31 & 0.7 & 0.215\\
\hline
 8.56$\times10^6$ & 1.74 & 18.9 & 9.14 & 0.7 & 0.536\\
\hline
 9.59$\times10^6$ & 1.65 & 16.8 & 9.12 & 0.7 & 0.628\\
\hline
 1.01$\times10^7$ & 1.64 & 16.6 & 9.11 & 0.69 & 0.640\\
\hline
 1.53$\times10^7$ & 1.63 & 16.4 & 9.12 & 0.69 & 0.652\\
\hline
 2.26$\times10^7$ & 1.63 & 16.4 & 9.12 & 0.69 & 0.652\\
\hline
 3.78$\times10^7$ & 1.64 & 16.5 & 9.10 & 0.69 & 0.640\\
\hline
 5.07$\times10^7$ & 1.64 & 16.5 & 9.08 & 0.68 & 0.640\\
\hline
 6.92$\times10^7$ & 1.66 & 16.6 & 9.05 & 0.67 & 0.62\\
\hline
 9.50$\times10^7$ & 1.68 & 16.7 & 9.02 & 0.66 & 0.6\\
\hline
 1.24$\times10^8$ & 1.70 & 16.9 & 8.98 & 0.64 & 0.57\\
\hline
 2.15$\times10^8$ & 1.77 & 17.4 & 8.86 & 0.60 & 0.51\\
\hline
 3.01$\times10^8$ & 1.85 & 17.9 & 8.74 & 0.55 & 0.45\\
\hline
 4.06$\times10^8$ & 1.96 & 18.6 & 8.57 & 0.48 & 0.37\\
\hline
 4.98$\times10^8$ & 2.08 & 19.2 & 8.39 & 0.42 & 0.31\\
\hline
 6.11$\times10^8$ & 2.26 & 20.0 & 8.12 & 0.33 & 0.24\\
\hline
 7.02$\times10^8$ & 2.45 & 20.5 & 7.85 & 0.25 & 0.19\\
\hline
 8.04$\times10^8$ & 2.75 & 20.9 & 7.45 &  0.15 & 0.14\\
\hline
 9.09$\times10^8$ & 2.76 &  25.5 & 7.81 & 2.6$\times10^{-3}$ & 0.13\\
\hline
 9.13$\times10^8$ & 2.78 &  27.8 & 7.95 & 2.2$\times10^{-4}$ & 0.13
\end{tabular}
\end{ruledtabular}
\end{table}


\begin{thebibliography}{99}

\bibitem{Hut} P. Hut, Astron.Astrophys. 99, 126 (1981).

\bibitem{Zahn1977} Zahn, Astron. Astrophys. 57, 383 (1977).

\bibitem{PressTeukolsky} W.H. Press, S.A. Teukolsky, Astrophys.J. 213, 183 (1977).

\bibitem{IvanovNovikov} P.B. Ivanov, I.D. Novikov, Astrophysic.J. 549, 467 (2001).

\bibitem{Rasio} F.A. Rasio, et.al. Astrophys.J. 470, 1187 (1996).

\bibitem{Penev} K. Penev, et.al. Astrophys.J. 751, 96 (2012).

\bibitem{BMathis} E. Bolmont, S. Mathis, ArXiv:1603.06268 [astro-ph].

\bibitem{IvanovPapChernov} P.B. Ivanov, J.C.B. Papaloizou, S.V. Chernov, MNRAS 432, 2339 (2013).

\bibitem{ChernovPapIvanov} S.V. Chernov, J.C.B. Papaloizou, P.B. Ivanov, MNRAS 434, 1079 (2013).

\bibitem{IvanovPapaloizou2004a} P.B. Ivanov, J.C.B. Papaloizou, MNRAS 347, 437 (2004).

\bibitem{IvanovPapaloizou2004b} P.B. Ivanov, J.C.B. Papaloizou, MNRAS 353, 1161 (2004).

\bibitem{IvanovPapaloizou2010} P.B. Ivanov, J.C.B. Papaloizou, MNRAS 407, 1609 (2010).

\bibitem{PapaloizouIvanov2010} J.C.B. Papaloizou, P.B. Ivanov, MNRAS 407, 1631 (2010).

\bibitem{LMathis} A.F. lanza, S. Mathis, ArXiv:1606.08623 [astro-ph].

\bibitem{Winn} J. Winn, D. Fabrycky, Annual Rev. Astron. Astrophys. 53, 409 (2015).

\bibitem{Ogilvie} G. Ogilvie, Annual Rev. Astron. Astrophys. 52, 171 (2014).

\bibitem{Brucalassi} A. Brucalassi, et.al. Astron. Astrophys. 561, L9 (2014).

\bibitem{Weinberg} N.N. Weinberg, et.al. Astrophys.J. 751, 136 (2012).

\bibitem{Essick} R. Essick, N.N. Weinberg, Astrophys.J. 816, 21 (2016).

\bibitem{PapaloizouIvanov2005} J.C.B. Papaloizou, P.B. Ivanov, MNRAS 364, L66 (2005).

\bibitem{IvanovPapaloizou2007} P.B. Ivanov, J.C.B. Papaloizou, MNRAS 376, 682 (2007).

\bibitem{Cowling} T.G. Cowling, MNRAS 101,367 (1941).

\bibitem{Zahn1970} Zahn, Astron. Astrophys. 4, 452 (1970).

\bibitem{Rocca} A. Rocca, Astron. Astrophys. 175, 81 (1987).

\bibitem{Zahn1975} Zahn, Astron. Astrophys. 41, 329 (1975).

\bibitem{Dalsgaard} J. Christensen-Dalsgaard, Lecture notes on stellar oscillations, 4th edn. (1998).

\bibitem{Brassard} P. Brassard, G. Fontaine, F. Wesemael, S.D. Kawaler, M. Tassoul, Astrophys. J. 367, 601 (1991).

\bibitem{paxton2011} B. Paxton, et.al. Astrophys.J.Supp. 192, 3 (2011).

\bibitem{paxton2013} B. Paxton, et.al. Astrophys.J.Supp. 208, 4 (2013).

\bibitem{paxton2015} B. Paxton, et.al. Astrophys.J.Supp. 220, 15 (2015).

\bibitem{Steffen} M. Steffen, Astron. Astrophys. 239, 443 (1990).



\end{thebibliography}
\end{document}